\documentclass[prl,twocolumn]{revtex4}
\usepackage{graphicx}% Include figure files
\usepackage{dcolumn}% Align table columns on decimal point
\usepackage{amsmath,amsbsy,amssymb,graphics}
\usepackage{bm}% bold math

\begin{document}

\preprint{APS/123-QED}

\title{Itinerant Antiferromagnetism in Infinite Dimensional Kondo Lattice}

\author{Shintaro Hoshino, Junya Otsuki, and Yoshio Kuramoto}
\affiliation{Department of Physics, Tohoku University, Sendai 980-8578}

\date{\today}

\begin{abstract}
Highly accurate numerical results for single-particle spectrum and order parameter are obtained for the  magnetically ordered Kondo lattice by means of the dynamical mean-field theory combined with the continuous-time quantum Monte Carlo method. Hybridized energy bands involving local spins are identified in the N\'{e}el state as a hallmark of itinerant antiferromagnetism. At the boundary of the reduced Brillouin zone, the two-fold degeneracy remains in spite of the doubled unit cell. This degeneracy results if the molecular field felt by localized spins has identical magnitude and reversed direction with that of conduction electrons. The persistent Kondo effect is responsible for the behavior. The antiferromagnetic quantum transition occurs inside the itinerant regime, and does not accompany the itinerant-localized transition.
\end{abstract}

\pacs{Valid PACS appear here}% PACS, the Physics and Astronomy
                             % Classification Scheme.
\maketitle

\newcommand{\diff}{\mathrm{d}}
\newcommand{\imag}{\mathrm{Im}\,}
\newcommand{\real}{\mathrm{Re}\,}
\newcommand{\trace}{\mathrm{Tr}\,}
\newcommand{\imu}{\mathrm{i}}

%.introduction
Distinction between itinerant and localized characters of strongly correlated electrons has been one of the most fundamental issues in condensed-matter physics.
The Kondo effect plays a central role in this problem because even a localized spin may acquire itinerant character by coupling with conduction electrons, and may form a Fermi liquid.
Some of recent experiment 
suggest  that 
quantum phase transition between magnetic and paramagnetic ground states
accompanies a change between localized and itinerant character of electrons,
with non-Fermi liquid behavior in the vicinity of the transition \cite{gegenwart}.
Another experiment using the de Haas-van Alphen effect has probed the change of the Fermi surface as the external pressure drives such systems as CeIn$_3$ and CeRhIn$_5$ across the magnetic transition \cite{onuki}. 
On the other hand, recent photoemission experiment for 
CeRu$_2$Si$_2$ and 
CeRu$_2$Si$_{2-x}$Ge$_x$ indicates that the Fermi surfaces of both are essentially the same, which 
involve $f$ electrons and are referred to as
the large Fermi surface.
The experiment is performed at temperatures above the N\'{e}el transition 
of CeRu$_2$Si$_{2-x}$Ge$_x$ \cite{okane},
while CeRu$_2$Si$_2$ remains paramagnetic down at least to 0.1 K.
Thus the antiferromagnetism in CeRu$_2$Si$_{2-x}$Ge$_x$ seems to be itinerant.
Consistent understanding of this variety of phenomena is still lacking, and accurate theoretical analysis is necessary to understand how the magnetic order is related to the change 
from itinerant to localized characters of electrons.

In this paper we show that the antiferromagnetism in the Kondo lattice model (KLM)
occurs within the itinerant regime.
The KLM is the simplest system that is capable of describing both itinerant and localized characters of electrons, and is given by
\begin{eqnarray}
{\cal H} = \sum_{\bm{k}\sigma} \varepsilon_{\bm{k}} c_{\bm{k}\sigma}^\dagger c_{\bm{k}\sigma} + 2J \sum _{i} \bm{S}_i \cdot \bm{s}_{{\rm c}i}, \label{eqn_klm}
\end{eqnarray}
where the first term represents the kinetic energy of conduction electrons,
$\bm{S}_i$ is the localized spin at site $i$, and 
$\bm{s}_{{\rm c}i}$
denotes the conduction-electron spin at the same site. 
In terms of creation 
$c_{i \alpha}^\dagger$ and annihilation $c_{i \alpha}$ operators 
of conduction electrons at site $i$ with spin $\alpha$, we obtain
$
\bm{s}_{{\rm c}i}
 = \frac{1}{2}\sum_{\alpha \beta} c_{i \alpha}^\dagger  \bm{\sigma}_{\alpha \beta}  c_{i \beta}
$.
The KLM has been investigated beyond the mean field theory in 
one- \cite{tsunetsugu2,shibata} and two- 
\cite{assaad,capponi, martin, watanabe, lanata}
dimensional systems.
We approach the KLM from infinite dimensions using the dynamical mean-field theory (DMFT)  
to allow for the N\'{e}el state at finite temperatures.
We use the continuous-time quantum Monte Carlo method (CT-QMC) \cite{rubstov, werner, otsuki3} as the impurity solver,
and focus on the case of 
unit number $n _{\rm c} = 1$ of conduction electrons per site.
In contast with the metallic antiferromagnetism with $n _{\rm c} \neq 1$, where 
an incommensurate order may exist,
we can safely assume the simple staggered order because of the nesting condition for the conduction band in the hypercubic lattice.
Hence the results in this paper are exact in infinite dimensions except for statistical errors. 

As the half-filled limit is approached from $n _{\rm c} \neq 1$, the large Fermi surface tends to the zone boundary of the paramagnetic phase, while the small Fermi surface involves half of the Brillouin zone volume.
Provided that the ground state has no discontinuity in the zero-doping limit,
the limiting location of the Fermi surface should be reflected in the location of the energy gap in the half-filled case.
In the localized antiferromagnetism, 
the gap opens at the boundary of the new Brillouin zone
 since conduction electrons feel the staggered internal field.
In itinerant magnetism, on the other hand, 
we show in this paper that the energy gap is located 
in the center of the new Brillouin zone.
The zone-center location is due to emergence of
energy bands of magnetic electrons.
Hence location of the energy gap distinguishes between itinerant and localized behaviors.
Note that such distinction does not apply to the single band model such as the Hubbard model where the energy gap in the half-filled limit always occurs in the boundary of the Brillouin zone.
In this case 
the character of electrons changes continuously from the itinerant limit to the localized one as 
the Coulomb repulsion increases relative to the band width.

We take the bare
density of states
\begin{equation}
\rho_{\rm c} (\omega) = \sqrt{2/\pi} \exp ({- 2 \omega ^2 })
\end{equation}
which corresponds to
infinite-dimensional hypercubic lattice.
We have taken the band width $D=1$ as the unit of energy.
For the Gaussian density of states,  the Kondo temperature $T_{\rm K}$ is defined by
\begin{equation}
\rho_{\rm c}(0) T_{\rm K} = \frac{e^{-\gamma /2}}{\sqrt{\pi}} \exp \left( -\frac{1}{2\rho_{\rm c}(0) J} \right),
\label{T_K}
\end{equation}
where $e^{-\gamma /2}/\sqrt{\pi} \sim 0.42$ with 
$\gamma \simeq 0.577$ being the Euler constant.
This expression of $T_{\rm K}$ corresponds to divergence of effective exchange in the lowest-order
scaling equation.

The hypercubic lattice has the nesting property with the wave vector $\bm{Q}=(\pi, \pi, \cdots)$ at half filling, which favors the staggered order.
In the two-sublattice formalism, the Green function 
$\bm{G}_{\bm{k}\sigma} (z)$ 
of conduction electrons 
is a $2\times 2$ matrix 
where $z$ is a complex energy, and
a wave vector $\bm{k}$ belongs to the reduced Brillouin zone \cite{georges}.
In the DMFT, 
the wave vector enters only through $\varepsilon _{\bm{k}}$.
Therefore we introduce the notation $\kappa = \varepsilon _{\bm{k}}$, and regard $\kappa$ as if it represents a wave number.
The spectral function $A_\sigma (\kappa , \omega)$ 
can be calculated from the matrix Green function 
$\bm{G}_\sigma (\kappa, z)$
as
\begin{align}
A_\sigma (\kappa , \omega) = - \imag \left[ \trace \bm{G}_\sigma (\kappa, \omega + \imu \delta) \right] /\pi.
\label{eqn_a_k_omega}
\end{align}
Then the renormalized density of states  $\rho _{\sigma} (\omega)$ is given by
$\rho _{\sigma} (\omega) = 2N^{-1} \sum_{\bm{k}} A_\sigma (\varepsilon _{\bm{k}} , \omega)$
\label{rho_sigma}
where
the summation runs over the reduced Brillouin zone with $N/2$ points.
In the paramagnetic state, the reduced number of $\bm{k}$ is compensated by the trace over $\bm{G}_{\bm{k}\sigma} (z)$ to give the same $\rho _{\sigma} (\omega)$ 
as derived by use of the original Brillouin zone.
Even in the N\'{e}el state, 
$\rho _{\sigma} (\omega)$ does not depend on $\sigma$ because of summation over sublattices.  
%Of course we can also derive the spin-dependent density of states for each sublattice.

\begin{figure}[b]
\begin{center}
\includegraphics[width=0.4\textwidth]{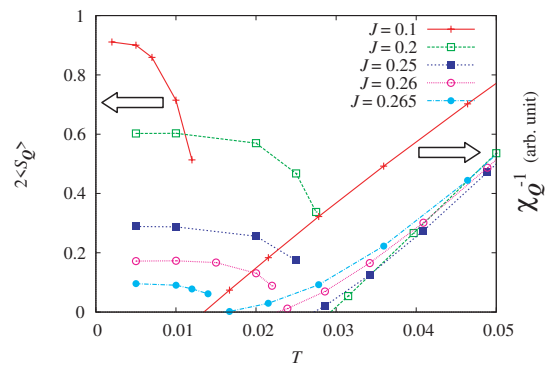}
\caption{(color online) 
Staggered spin polarization 
2$\langle S ^z _{\bm{Q}} \rangle$ (left scale)
as a function of temperature for different values of $J$.  
Also shown is  
the inverse staggered susceptibility $\chi _{\bm{Q}} ^{-1}$ (right scale),
which goes to zero at $T_{\rm N}$
}
\label{fig_mom}
\end{center}
\end{figure}

Let us first discuss the magnitude of the staggered moment given by
$
2\langle S ^z _{\bm{Q}} \rangle = 
\langle S ^z _{\rm A} \rangle -\langle S ^z _{\rm B} \rangle
$,
where we choose the positive polarization for the A sublattice.
Figure \ref{fig_mom} shows the temperature dependence of 
the staggered moment,
which should vanish at the N\'{e}el temperature $T_{\rm N}$.
On the other hand, the staggered 
magnetic susceptibility $\chi _{\bm{Q}}$ in the paramagnetic state should diverge
as the temperature is lowered toward $T_{\rm N}$.
Hence we also plot $\chi _{\bm{Q}}^{-1}$ 
calculated in ref.\cite{otsuki1}.
It is found that the estimates of $T_{\rm N}$ by 
$\langle S ^z _{\bm{Q}} \rangle$ and by $\chi _{\bm{Q}}^{-1}$ are consistent with each other.
However, calculation of 
$\langle S ^z _{\bm{Q}} \rangle$ becomes increasingly difficult as $T$ is approached to $T_{\rm N}$.
The results shown in FIG.\ref{fig_mom} are restricted to the 
temperature range where we could obtain reliable values.
With $J=0.1$, the localized spins are almost fully polarized in the ground state.
%The temperature dependence can be fitted very well by the mean-field formula, provided the saturated magnetization at $T=0$ is given.
As $J$ increases, $\langle S ^z _{\bm{Q}} \rangle$ 
decreases by the Kondo effect, which eventually suppresses the antiferromagnetism
at $J=J_{\rm c}\simeq 0.27$ down to $T=0$.
%With $J\gtrsim 0.2$,  the mean-field theory does no longer provide a good fit for the temperature dependence.

\begin{figure}%[tb]
\begin{center}
\includegraphics[width=0.4\textwidth]{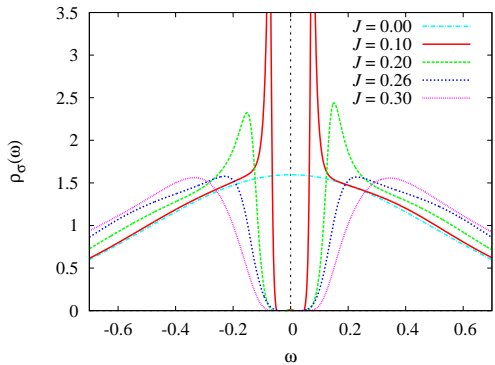}
\caption{(color online) $J$ dependence of the density of states for the conduction electron and with $T \sim 0$.}
\label{fig_dos_g}
\end{center}
\end{figure}

We next discuss the density of states near $T=0$, which is derived
using the Pad$\acute {\rm e}$ approximation
for analytic continuation from imaginary Matsubara frequencies $\imu \epsilon_n$ to the real ones.
Since the Monte Carlo data are obtained accurately in the imaginary time domain, the Pad$\acute {\rm e}$ approximation works well in the CT-QMC method\cite{otsuki3}.
Figure \ref{fig_dos_g} shows $\rho _{\sigma} (\omega)$ 
for different values of $J$ near the ground state.
Namely, we take the temperature where the density of states does not vary 
much
when $T$ decreases further.
  For example, we take $T=0.01$ for $J=0.3$, but 
$T=0.002$ for $J=0.1$.
The present method has no difficulty to go to such low temperatures with high accuracy.
In the case of $J=0.3$ with the paramagnetic ground state, the density of states has a gap caused by the Kondo effect.  The state
is often called the Kondo insulator.
In the case of $J=0.26$, the ground state is antiferromagnetic as seen from FIG.\ref{fig_mom}.
The density of states in the ordered phase is almost the same as that with $J=0.3$.
Namely the density of states does not depend much on whether the ground state is paramagnetic or antiferromagnetic as far as $J$ is close to the critical value $J_{\rm c}$.

When $J$ is smaller, 
sharp peaks develop at both edges of the gap.
The origin is
understood as follows: 
Putting the self-energy of conduction electrons 
as the staggered potential $\pm h$ by the N\'{e}el order, 
we obtain
\begin{equation}
\trace \bm{G}_\sigma (\kappa, z)  =  \frac{1}{z + \sqrt{\kappa^2+h^2}} + \frac{1}{z - \sqrt{\kappa^2+h^2}}.
\label{staggered}
\end{equation}
Then the density of states is given by
\begin{equation}
\rho_\sigma (\omega) = 
\frac{2|\omega|}{\sqrt{\omega^2-h^2}}
\ \rho_c (\sqrt{\omega^2-h^2}),
\end{equation}
with a square-root divergence at both edges $\omega=\pm h$. 
For $\omega^2<h^2$, we obtain $\rho_\sigma (\omega)  = 0$.
This form of $\rho_\sigma (\omega)$
roughly explains the peak structure with $J=0.1$ in FIG.\ref{fig_dos_g}.
In the numerical result,
the residual Kondo effect actually suppresses the divergence in  $\rho_\sigma (\omega)$.

\begin{figure}%[tb]
\begin{center}
\includegraphics[width=0.4\textwidth]{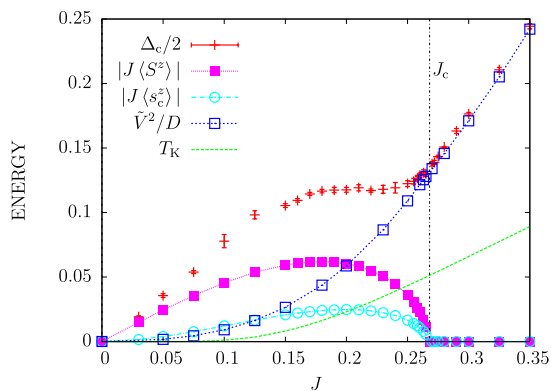}
\caption{(color online)
$J$ dependence of the energy gap $\Delta _{\rm c} /2 $.
For comparison, also plotted are $J \langle S^z \rangle$, $J \langle s_{\rm c}^z \rangle$,
$S^z$ and $s_{\rm c}^z$ are the localized and conduction spins at a local site, respectively.
${\tilde V}^2 / D$ represents the effective hybridization to be explained later,
 and $T_{\rm K}$ is the Kondo temperature defined by Eq.(\ref{T_K}).
}
\label{fig_gap}
\end{center}
\end{figure}

On the other hand, the density of states for larger $J$ does not show a clear threshold.
The Gaussian tail in the bare density of states $\rho _{\rm c} (\omega)$
causes a tiny but finite magnitude of $\rho _{\sigma} (\omega)$ 
within the apparent energy gap.
Therefore we introduce the characteristic value 
$\Delta _{\rm c}$ of the energy gap as giving 
the half-maximum value of the peak in the density of states.
Figure \ref{fig_gap} shows $\Delta _{\rm c}/2$ as a function of $J$.
The error bars have been estimated from 5 bins of data.
It is clear that $\Delta _{\rm c}$ changes continuously at $J=J_{\rm c}$.
This shows that both Kondo effect and the staggered internal field are contributing to $\Delta _{\rm c}$.
For $J\lesssim 0.1$, on the other hand, $\Delta _{\rm c}/2$ is almost
proportional to $J$.
This behavior shows that the gap is mainly determined by the staggered field
$J\langle S^z \rangle$
as shown also in FIG. \ref{fig_gap}
.

\begin{figure}%[b]
\begin{center}
\includegraphics[width=0.45\textwidth]{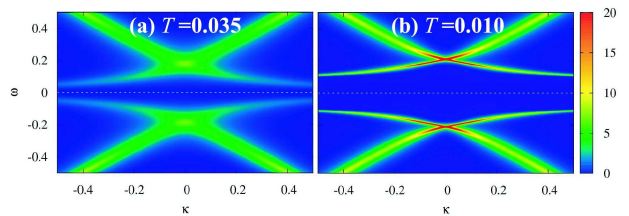}
\caption{(color online) 
Single-particle spectrum with $J=0.2$ in (a) paramagnetic phase at $T=0.035$ and (b) antiferromagnetic phase at $T=0.010$.}
\label{fig_spect1}
\end{center}
\end{figure}

The details of the itinerancy are seen in the single-partcle spectral function $A_{\sigma} (\kappa , \omega) $.
Figure \ref{fig_spect1} shows the spectrum in
the case $J=0.2$ where the Kondo effect is significant.
In both paramagnetic and ordered phases, the spectrum consists of four bands
in the reduced Brillouin zone.
In the paramagnetic phase shown in the left panel,  
the new bands are ascribed to 
``hybridization bands" caused by the Kondo effect \cite{otsuki_lett}.
We note that there is no real hybridization between the $f$-electron and the conduction electron because the $f$-electron 
does not have the charge degrees of freedom
in the KLM.
In the Brillouin zone of the paramagnetic state, the energy gap is indirect from the zone boundary to the zone center, both of which correspond to $\kappa\rightarrow \pm\infty$ in FIG.\ref{fig_spect1}. 
In the right panel of FIG.\ref{fig_spect1}(b), 
these hybridization bands are clearly seen even in the antiferromagnetic phase.
Hence we classify the antiferromagnetism in this regime as itinerant.

\begin{figure}%[tb]
\begin{center}
\includegraphics[width=0.45\textwidth]{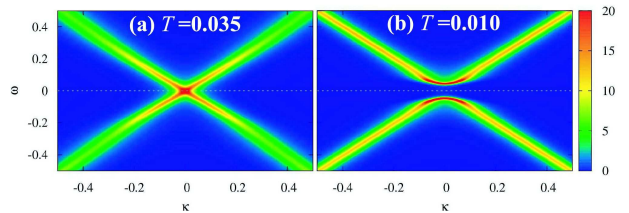}
\caption{(color online) $A_\sigma (\kappa, \omega)$ with ferromagnetic interaction $J=-0.2$ in (a) paramagnetic phase at $T=0.035$ and (b) antiferromagnetic phase at $T=0.010$.}
\label{fig_spect2}
\end{center}
\end{figure}

Let us now present the spectrum in the ferromagnetic KLM for comparison, where 
each site forms $S=1$ state with antiferromagnetic intersite interaction.
The Kondo effect is absent in the case of
$J<0$. 
We can apply the CT-QMC method also to ferromagnetic $J$ as noted in ref. \cite{hoshino}.
Figure \ref{fig_spect2} shows the spectrum with the same $T$ as in FIG.\ref{fig_spect1}.
In the paramagnetic state shown in FIG.\ref{fig_spect2}(a), the spectrum is almost the same as the original conduction band.
On the other hand, in the antiferromagnetic phase shown in FIG.\ref{fig_spect2}(b), the spectrum shows the clear gap structure.
These behaviors are explained well by Eq.(\ref{staggered}) with $h=0$ in FIG.\ref{fig_spect2}(a), while $h\neq0$ in (b).
In the case of ferromagnetic $J$, 
the number of bands is two 
in contrast to the case 
with $J>0$ with four bands.

Thus, the itinerant or localized behavior can be distinguished by the number of energy bands in the reduced Brillouin zone.
Moreover the location of the energy gap is at the zone boundary in the localized case, but at the zone center in the itinerant case.
Note that $\kappa = \pm \infty$ correspond to the lower and upper edges of the conduction band in infinite dimensions.
Both edges come to the center in the reduced Brillouin zone.

We note another characteristic feature in FIG.\ref{fig_spect1}(b) 
that the degeneracy at $\kappa = 0$ remains in the antiferromagnetic phase.
A toy model helps to understand the origin of the degeneracy.
Let us consider a non-interacting periodic Anderson model under staggered field as
\begin{eqnarray}
{\cal H}_{\rm PAM} &=& \sum_{\bm{k}\sigma} \varepsilon_{\bm{k}} c_{\bm{k}\sigma}^\dagger c_{\bm{k}\sigma} + V\sum _{i\sigma} (f_{i\sigma}^\dagger c_{i\sigma} + {\rm h.c.}) \nonumber\\
&& + 2\sum_{i}(h_i s_{{\rm c}i}^z + H_i S_i^z ) , \label{eqn_toy}
\end{eqnarray}
where 
$f_{i\sigma}(f_{i\sigma}^\dagger)$ is the annihilation (creation) operator of an $f$-electron at the $i$ site with
the hybridization $V$.
The staggered fields represent the molecular field associated with the antiferromagnetism.
Then we choose  
$h_i = \pm h$ and
$H_i = \pm H$, where the A (B) sublattice has the positive (negative) sign.

There are 
four branches associated with c and f degrees of freedom for the electrons, 
as well as presence of A and B sublattices.
At $\kappa = 0$, we obtain the energies
\begin{eqnarray}
E(\kappa = 0) = \pm \frac{1}{2} \left[ (h+H) \pm \sqrt{(h-H)^2 + 4V^2} \right] .\label{eqn_toy_ene}
\end{eqnarray}
If the relation $h = -H$ holds, $E(\kappa = 0)$ has only two distinct values both of which are doubly degenerate.
We interpret the degeneracy in 
FIG.\ref{fig_spect1}(b) as caused by the relation $h = -H$ for the molecular field.
We call such situation ``quasi-local compensation".
Note that
the magnitudes of the polarization for the conduction spin $J\langle s_{\rm c}^z \rangle$ and localized spin $J\langle S^z \rangle$ are much different as shown in FIG.\ref{fig_gap}.
Hence the actual energy level is determined not by the local internal field, but by a long-range field involving remote conduction electrons.
The compensation is reminiscent of the spatially extended Kondo singlet with small $J$.

Let us identify the energy scale in the N\'{e}el state from
the self energy.
For sublattice $\alpha=\pm 1$ and spin $\sigma=\pm 1$,
$\Sigma _{\alpha \sigma} (z)$ 
is expanded as
\begin{eqnarray}
\Sigma _{\alpha \sigma} (z) = \alpha \sigma h + \frac{\tilde V^2}{z} + {O} \left( \frac{1}{z^2} \right) , \label{eqn_self_toy}
\end{eqnarray}
where $\tilde V$ is the effective hybridization.
The coefficient $\tilde V^2$ of $1/z$ corresponds to the jump at $\tau = 0$ in the imaginary time domain.
We extract numerically the coefficient ${\tilde V}^2$ 
of $1/ \imu \varepsilon _n$ in the self energy.
Figure \ref{fig_gap} shows the result for 
${\tilde V}^2/D \ (=\tilde V^2)$ as a function of $J$.
Note that 
the indirect energy gap in the toy model (\ref{eqn_toy}) is given by $V^2/D$.
The agreement between ${\tilde V}^2 / D$ and $\Delta _{\rm c} /2$ in the paramagnetic phase is excellent.  
However, this consistency should not be taken too seriously 
because $\Delta _{\rm c}$ depends on the definition of the gap.
We emphasize that ${\tilde V}^2 / D$ shows no anomaly
across the phase transition to 
the antiferromagnetic phase.
It also shows good proportionality to the Kondo temperature 
as ${\tilde V}^2 / D \simeq 2.6 T_{\rm K}$.

In the region $0<J\lesssim 0.1$, the N\'{e}el temperature 
$T_{\rm N}$ is much larger than $T_{\rm K}$  
as seen from FIGS.\ref{fig_mom} and \ref{fig_gap}. 
The electronic state at the transition
has a localized character since the Kondo effect is negligible at $T_{\rm N}$.   Namely
there is no hybridized band, and the energy gap occurs at the boundary of the reduced Brilluoin zone.  
However, 
we have checked that two almost flat bands appear newly below $T_{\rm K}$
even with $J=0.05$.
Hence, the crossover from localized behavior to the itinerant one occurs inside the N\'{e}el ordered state.

In summary, 
we have derived single-particle spectrum and the temperature-dependent order parameter
in the infinite dimensional KLM allowing for the N\'{e}el order.
The high numerical accuracy has made it possible to find
the quasi-local compensation between the conduction and localized spins indicating the persistent tendency toward the Kondo singlet at each site.
The effective hybridization energy has no anomaly across the quantum phase transition, and scales well with the impurity Kondo temperature $T_{\rm K}$.
Hence the quantum transition into antiferromagnetism occurs within the itinerant regime, and does not involve itinerant-localized transition.

\end{document}